# Relations between photoionization cross sections and photon radius

Shan-Liang Liu

Shandong Key Laboratory of Optical Communication Science and Technology, School of Physical Science and Information Engineering, Liaocheng University, Shandong 252059, People's Republic of China

The relations between photoionization cross sections and photon radius are obtained on basis of quantum mechanics and the particle-like properties of a photon. The photoionization cross sections of H atom and H-like ions, He atom and He like ions, alkali metal atoms, and Rydberg atoms are calculated using the relations. The calculation results are found to be good agreement with the known experimental data. The results show that the photoionization cross section is always smaller than the cross section of the photon to ionize the atom or ion and can be expressed as the product of the cross section of the photon and the probability that electron meets with the photon. These provide the intuitive understanding for the photoionization phenomena and open a new avenue of research on interaction between a photon and an atom or ion.

## 1. Introduction

An atom or ion is ionized by a single photon whose energy $h\nu$ is higher than the ionization energy $U_i$, and the kinetic energy of the photoelectrons is given by $h\nu - U_i$, which is the most obviously particle-like characteristic of a photon. Recently, it has been shown that a photon has the shape of a cylinder with length of $\lambda/2$ and its radius is proportional to square root of the wavelength [1]. However, the photoionization cross section is usually calculated by the dipole approximation which results from the combination of quantum mechanics and wave theory of electromagnetic fields [2-4]. There is not intuitive understanding for some phenomena that the photo-ionization cross section monotonously decreases with increase of the photon energy; the electrons in inmost shell in the atoms is ionized easier than those in the outer shell for the same photon energy which can ionize them, and so on. The dipole approximation is theoretically no longer valid in some cases, but the calculations that make the approximation are in agreement with the experimental data [5, 6]. To reconcile this apparent inconsistency, one may argue that the photo-ionization process occurs close to the nucleus in the atom [7, 8]. The measured values of the photo-ionization cross sections for potassium (K) atoms in the 5d and 7s excited states are much larger than the theoretical values [9,10], whereas the measured values of the cross section for the Rydberg rubidium (Rb) atom in the nd states are lower by about a factor of 4 than the theoretical values [11]. The striking differences between the theoretical calculations and experimental data are still to be further investigated.

The relations between photoionization cross sections and photon radius are obtained here on basis of quantum mechanics and the particle-like properties of a photon. The calculations from the relations are found to be in good agreement with the known experimental data. These provide the intuitive understanding for the photoionization phenomena and open a new avenue of research on interaction between a photon and an atom or ion.

## 2. Photoionization of atoms and ions

### 2.1. Photoionization of H and H-like ions

Radius of a photon is proportional to square root of the wavelength and is given by [1],

$$\rho_0 = \frac{2\sqrt{2r_e\lambda}}{(\sqrt{2}-1)\pi}, \qquad (1)$$

where $r_e = e^2/(4\pi\varepsilon_0 m_0 c^2)$ is the classical radius of an electron. Since the electron acts with the photon only if an electron is in the electromagnetic fields of a photon, the cross section of a photon $S_0 = \pi\rho_0^2$ is the interaction cross section between an electron and a photon. There is only one electron in an H atom or H-like ion, the ionization energy $U_i = Z^2 U_H$ at the ground state where $Z$ is the proton number in the atom or ion and $U_H = 13.6$ eV is the ionization energy of the H atom. Substituting $\lambda = hc/U_i$ into Eq. (1) gives the maximum radius of the photon to ionize the atom or ions

$$\rho_{\max} = 0.66 r_1, \qquad (2)$$

where $r_1 = a_0/Z$, $a_0$ is the Bohr radius. Equation (2) reveals that an H atom or H-like ion can be ionized only by the photon whose radius is shorter than $r_1$.

The radial wavefunction of the electron in H atom or H-like ion at the ground state is

$$R_{10} = 2(1/r_1)^{3/2} \exp(-r/r_1). \qquad (3)$$

The probability density $|R_{10}|^2$ of the electron exponentially decreases with increase of $r$ and is at

$r=r_1$ only $1/e^2$ of the maximum at $r=0$. The probability of the electron in the cylinder region $V_0$ with radius of $\rho_0$ has the maximum at $\rho=0$

$$P_0 = \int_0^\infty \int_0^{\rho_0} R_{10}^2 \rho d\rho dz, \qquad (4)$$

where $r=(\rho^2+z^2)^{1/2}$, $\rho$ denotes the distance of the nucleus from the photon trajectory, and $z$ is in the motion direction of photon. If the probability the electron meets with the photon in the cylinder region is negligible at $\rho>\rho_0$ and can be expressed as $(\rho_0/r_1)^2$ at $\rho<\rho_0$, the probability that the electron meets with the photon $P=P_0(\rho_0/r_1)^2$ according to the statistical law, and the photoionization cross section of the H atom or H-like ion is given by

$$\sigma = \pi \rho_0^2 P_0 (\rho_0/r_1)^2, \qquad (5)$$

which monotonously increases with $\rho_0$ or decrease of the photon energy.

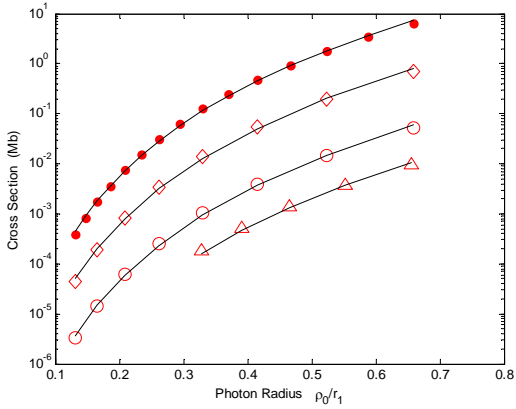

FIG. 1. Photoionization cross sections of H atom and H-like ions as function of photon radius. The solid curves are plotted by Eq. (5), the experimental data: ● for H, ◊ for $Li^{+2}$, ○ for $Na^{+10}$; △ for $Fe^{+25}$) are taken from Ref. [12].

The cross sections of H atom and all H-like ions of the Opacity Project (OP) elements are calculated using Eq. (5). Figure 1 shows the cross sections of the H atom and H-like ions as function of the photon radius in unit of $r_1$, where $r_1$ varies with Z and has different values for different ions. The calculation results (solid curve) agree with the experimental data (signs) very well for $\rho_0>0.1r_1$, as shown in Fig. 1. The calculation results are in good agreement with the experimental data for the photon energy from the ionization threshold to 50 kev if the probability that the electron meets with the photon is given by

$$P = \left(\frac{\rho_0}{r_1}\right)^4 \left[1 - \exp\left(-\frac{12\rho_0}{r_1}\right)\right]. \qquad (6)$$

For $\rho_0>0.1r_1$, $P\approx(\rho_0/r_1)^4$, being in good agreement with $P_0=(\rho_0/r_1)^2$ obtained from Eq. (4), and the cross section is given by

$$\sigma = \pi \rho_0^2 (\rho_0/r_1)^4, \qquad (7)$$

which is proportional to $Z^4$ and $\rho_0^6$ or $\lambda^3$. For $\rho_0<<r_1$, $\sigma=12S_0(\rho_0/r_1)^5$ is proportional to $Z^5$ and $\rho_0^7$ or $\lambda^{7/2}$. It follows that the photoionization cross section rapidly decreases with the photon radius or with increase of the photon energy for some ion and rapidly increases with the atomic number Z for the given photon energy because the orbital radius $r_1$ is inverse proportional to Z.

**2.2. Photoionization of He and He-like ions**

According to the Bohr's theory, the threshold energy of the double photoionization (DPI) in the He atom or He-like ion at the ground state

$$U_d = 2Z_1^2 U_H. \qquad (8)$$

Two electrons have the same orbital radius

$$r_1 = a_0/Z_1, \qquad (9)$$

where $Z_1=Z-s$ is the effective atomic number, $s$ is known as the screening constant and arises from interaction between two electrons in the atom or ion. The parameters s and $r_1$ for all the He-like ions are calculated from Eqs. (8) and (9) using the measured values of $U_d$ in Ref [12]. The screening constant of the He-like ion at $Z<13$ has almost the same value of $s=0.3$ as that of He and obeys Slater's rule, but the screening constant slowly decreases with increase of Z for $Z>13$, and $s=0.2$ for $Fe^{+24}$ is obviously smaller than 0.3 from Slater's rule. It means that Slater' rule is no longer accurate for $Z>13$

The Substituting $\lambda=hc/U_1$ into Eq. (1) gives the maximum radius of the photon to ionize the atom or ions

$$\rho_{max} = 1.286\times10^{-10} U_1^{-1/2}, \qquad (10)$$

where $U_1$ is the ionization threshold of the He atom or He-like ions in the unit of eV. The maximum radius of the photon is obtained by substituting the measured values of $U_1$ into Eq. (10). The maximum ratio $\rho_{max}/r_1$ monotonously decreases from 0.83 to 0.67 as Z increases from 2 to 26 and is smaller than that of H atom or H-like ions. The larger Z is, the closer $\rho_{max}/r_1$ of the He-like ion is to that of the H-like ion. This shows that the effect of interaction between two electrons on their states becomes the weaker and weaker as Z increases; the He atom or He-like ion at the ground state can be ionized only by the photon whose radius is smaller than the orbital radius.

There are two identical electrons in the He atom or He-like ion, the wavefunction and $P_0$ have the same form as Eqs. (3) and (4) for each of them if the independent particle approximation is made, respectively, and the photoionization cross sections $\rho_0>0.1r_1$ is given by

$$\sigma = 2\pi \rho_0^2 (r_0/r_1)^4. \qquad (11)$$

The cross sections of He atom and all He-like ions of the Opacity Project (OP) elements are calculated using Eq. (11). The calculation results deviate from the experimental data for the He atom near the ionization threshold due to interaction between two electrons, especially. With increase of $Z$, the deviation as well as the interaction effect becomes smaller and smaller, the calculation results for $Fe^{+24}$ are in good agreement with the experimental data (Δ), as shown in Fig. 2.

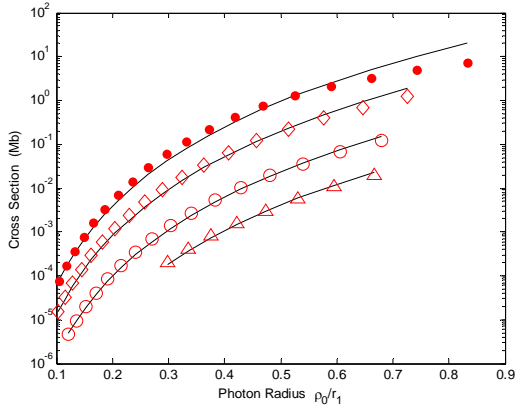

FIG. 2. Photoionization cross section of He and He-like as function of photon radius. The curves are the calculations results, the experimental data (● for He, ◊ for $Be^{+2}$, ○ for $Na^{+9}$; Δ for $Fe^{+24}$) are taken from Ref. [12].

### 2.3. Photoionization of alkali metal atoms

The probability that the valence electron at alkali metal atom occurs within the cylinder region $V_0$ has the maximum

$$P_0 = 4\pi \int_0^\infty \int_0^{\rho_0} |\psi_{nl0}|^2 \, \rho d\rho dz . \qquad (12)$$

where $\psi_{nlm} = R_{nl}(r)Y_{l0}(\cos\theta)$ is the normalized wave function, $Y_{l0}(\cos\theta)$ is the spherical harmonic function, and $R_{nl}(r)$ are Slater radial wavefunction of the valence electron [13,14], respectively. The Slater radial wavefunctions are given by

$$R_{nl} = N_{nl}\left[ r^{n-1}\exp(-r/r_n) + \sum_{j=1}^{n-1} A_{nlj} R_{jl} \right], \qquad (13)$$

$$N_{nl} = \left[ (2n)!(r_n/2)^{2n+1} - \sum_{j=l+1}^{n-1} A_{nlj}^2 \right], n > l, \qquad (14)$$

$$A_{nlj} = N_{jl}\left[ \sum_{k=1}^{j-1} A_{nlk} A_{jlk} - \frac{(n+j)!}{(r_j + r_n)^{n+j+1}} \right], \qquad (15)$$

$$r_n = \begin{cases} na_0/Z_n, & Z_n > 1 \\ n^* a_0, & Z_n = 1 \end{cases} \qquad (16)$$

where $Z_n$ is the effective atomic number and is determined by Slater's rule, $n^* = n - \delta_l$ is the effective quantum number, and $\delta_l$ is the quantum defect. For Li atom at the ground state, $r_1 = a_0/2.7$, the calculation results (black dotted curve) using Eq. (5) and the Slater wavefunction are agreement with experimental data (red dots) for $\rho_0 < r_1$ and obviously deviate from the experimental dada for $\rho_0 > r_1$, as shown in Fig. 3.

Since the probability that electron meets with the photon is equal to $P_0$ in the case of $\rho_0 > r_1$, instead of $P_0(\rho_0/r_1)^2$, and the photoionization cross section $\rho_0 > r_1$ is given by

$$\sigma = \pi \rho_0^2 P_0 . \qquad (17)$$

The calculation results (solid curves) for the Li and Na atoms at the ground state using Eqs. (12)-(17) for $\rho_0 > r_1$ are in agreement with experimental data except near the ionization threshold, as shown in Fig. 3. The differences between the theoretical calculations and the experimental data near the ionization threshold can be attributed to the core polarization [10]. The cross section increases with the photon radius for $\rho_0 > r_1$.

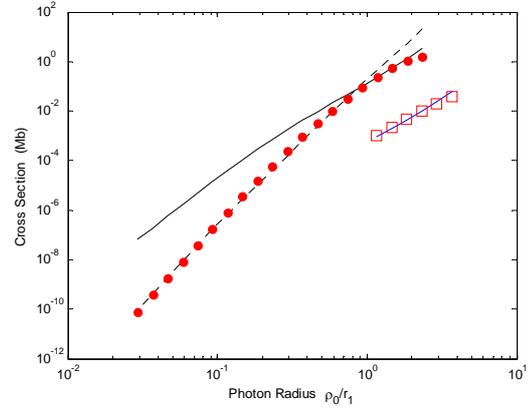

FIG. 3. Photoionization cross section of Li and Na atoms as function of photon radius. The solid curves are the calculations result from Eq. (17), the dotted curve from Eq. (5). The signs denote the experimental data of Li atom (●) and Na atom (□) from Ref. [12].

The photoionization cross sections of K atoms in the 4p, 5d, and 7s excited states as function of the photon radius are calculated using Eqs. (12)-(17) for $\rho_0 > r_1$. The calculated value of K atom in the 4p excited state for the 355-nm photon is 8 Mb and agrees with the measured value of 7.2±1.1 Mb [9]. The calculated value of K atom in the 7s excited state for the 660.6-nm photon is 0.67 Mb and agrees with the measured value of 0.61±0.09 Mb [9], which are larger by two orders of magnitude than the previous calculation [10]. However, the calculated value of K atom in the 5d excited state for the 662.6-nm photon is 10 Mb, about the twice the previous calculation value [10], and is obviously smaller than the measured value of 28.9±4.3 Mb [9].

The probability density of valence electron in K atoms in the ns, np, and nd states is calculated using the Slater wavefunction. Unlike the ns state, the probability density is zero at $r=0$ in the np and nd states, and so the electron and the photo-ionization process hardly occur near the nucleus.

The probability density (red plus) at $r=1.78a_0$ (the photon radius at 662.6 nm) is only about 1/3 of the maximum, as shown in Fig. 4 (a). The probability in the cylinder with the cross section $S_0$ between $\rho$ and $\rho+\rho_0$ in the 5d state slowly decreases with increase of $\rho$ and is still more than 1/3 of the maximum at $\rho=6a_0$, and the valence electron is in the region of $\rho>\rho_0$ with large probability, as shown in Fig. 4(b). Since $\rho_0<r_{max}$ at which the probability density has maximum, the probability that the electron meets with the photon in the region of $\rho>\rho_0$ can not be neglected, and the calculated values from Eq. (17) would be less than the true values.

If $S_{eff}$ represents the effective cross section of the atom or ion in which the electron must occur at any point with the same probability, the probability of the electron in the cylinder region with the cross section of $S_0$ is $S_0/S_{eff}$. The probability that photon is in the cylinder region during a photon goes through the atom or ion is also $S_0/S_{eff}$. Therefore, $S_{eff}$ can be considered as the interaction cross section between a photon and an atom or ion for $\rho_0<r_{max}$, the probability that the electron meets with the photon is $(S_0/S_{eff})^2$, and the photoionization cross section is given by

$$\sigma = S_{eff}(S_0/S_{eff})^2 = S_0^2/S_{eff}. \quad (18)$$

Equation (18) gives $\sigma=27$ Mb for K atom in the 5d state and the 662.6-nm photon if the effective radius $r_{eff}=5.7a_0$ is taken, which is in good agreement with the measured value. It is seen in Fig. 4 that $r_{eff}>r_{max}>\rho_0$, the probability density (blue circle) at $r=r_{eff}$ is obviously larger than that at $r=\rho_0$, and the averaged probability at $\rho=r_{eff}$ is about 1/3 of the maximum at $\rho=0$. For alkali metal atoms, $n^*>1$, $\rho_{max}=0.66n^*a_0$ is obviously shorter than the atom radius $n^{*2}a_0$.

### 3. Photoionization of Rydberg atoms

The radial wavefunction of Rydberg atoms can be given by [15,16]

$$R_{nl} = \frac{\sqrt{r_n^3 n_r!} r'^{l*} \exp(-r') L_{n_r}^\mu(r')}{\sqrt{(2n_r+\mu+1)\Gamma^3(n_r+\mu+1)}}, \quad (19)$$

where $r'=r/r_n$, $r_n=n^*a_0$, $\mu=2l^*+1$, $l^*=l-\delta_l>0$, $n_r=n-l-1$, and $L_{n_r}^\mu$ is the generalized Laguerre polynomial. The probability density distributions of valence electron in the nd states of Rydberg atoms are calculated using the wavefunction (19). The calculation results show that the maximum of probability density decreases with $\delta_l$ and with increase of n; $r_{max}$ hardly varies with n, but it increases with decrease of $\delta_l$. In the 16d state $r_{max}=a_0$ for the Rb atom ($\delta_l=1.345$) and $r_{max}=4.6a_0$ for the H atom ($\delta_l=0$), and $r_{max}=4a_0$ for K atom ($\delta_l=0.231$) in the 5d state is similar to that obtained using the Slater wavefunction, as shown in Fig. 5(a).

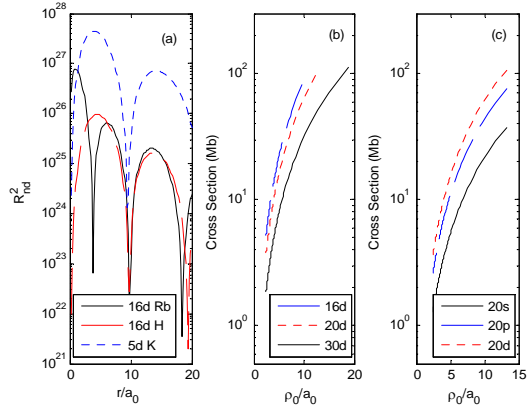

FIG. 5. (a) Probability density distributions of the different atoms in the nd states. The photoionization cross section as function of photon radius: (a) for Rydberg Rb atoms with different values of n and (c) for Rydberg H atoms with different values of $l$.

For Rydberg atom $n^*>>1$, $\rho_{max}=0.66r_n>>r_{max}$, and the photoionization cross section can be calculated using Eq. (17). The photoionization cross sections of Rydberg Rb atoms in the nd states are calculated using the relation (17) and the wavefunction (19) from the ionization threshold to 1 eV. The cross section decreases with increase of $n$ for the given photon energy and increase with the photon radius for the given value of $n$, as shown in Fig. 5(b). For the 10.6-μm photon, $\rho_0=7.1a_0$ is obviously larger than $r_{max}=a_0$, the calculations for the Rb atoms in the nd states ($16\leq n\leq 20$) give 32 Mb, 26 Mb, 21.1 Mb, 17.3 Mb, and 14.5 Mb, respectively. The measured values using the 10.6 μm laser in a magneto-optical trap were 39 Mb for 16d, 35 Mb for 17d, 32 Mb for 18d, and 30 Mb for 19d and 20d, respectively, the total measurement error can be estimated around 40% [11]. If the statistical errors are included, the calculation results can be considered to be in agreement with the experimental data.

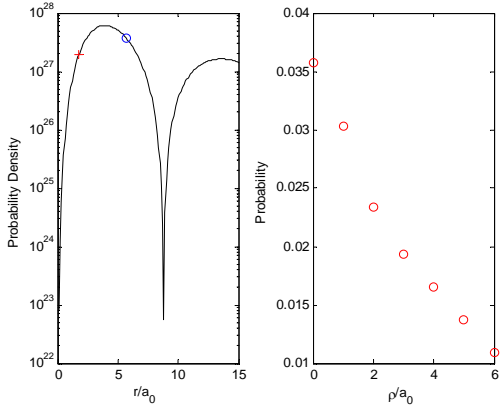

FIG. 4. (a) Probability density distribution of the valence electron in the 5d state of K and (b) the averaged probability in the cylinder with the cross section of $\pi\rho_0^2$ between $\rho$ and $\rho+\rho_0$.

The photoionization cross sections of Rydberg H atoms with different values of $l$ are calculated using the formula (17) and the wavefunctions (19). When the photon energy increases from the ionization threshold of 0.034 eV at $n$=20 to 1 eV, the photon radius decreases from $13.2a_0$ to $4.3a_0$, the cross section increases about 30 times, as shown in Fig.5(c). The cross section of $l$=1 (blue dashed curve) is about twice as large as that of $l$=0 (black solid curve) and slightly lower than that of $l$=2 (red dotted curve).

## 4. Conclusion

The relations between photoionization cross sections and photon radius are obtained on basis of quantum mechanics and the particle-like properties of a photon. The calculations from the relations are in good agreement with the known experimental data. The photoionization cross section is always smaller than the cross section of the photon to ionize the atom or ion and can be expressed as the product of the cross section of the photon and the probability that electron meets with the photon. The photonionization mainly occurs in the case of the nucleus in a line of the photon trajectory. The photoionizaton probability of the electron near the nucleus has the maximum in the ns state and is equal to zero in the np and nd states. The atom or ion can be ionized only by the photon whose radius is smaller than the atom or ion radius. The photo-ionization cross section rapidly decreases with the photon radius for the electron in the given state and rapidly increases with the atomic number Z for the given photon energy. The photoionization cross section of Rydberg atom decreases with increase of the principal quantum number for the given photon energy and increase with the photon radius for the given principal quantum number. **The maximum probability density of the valence electron in Rydberg atoms decreases with the quantum defect and with increase of the principal quantum number.** These provide the intuitive understanding for the photoionization phenomena and open a new avenue of research on the photon interaction with atoms and ions.